\documentclass[apjl]{emulateapj}

\usepackage{graphicx}
\usepackage{apjfonts}
\slugcomment{Accepted for The Astrophysical Journal Letters, 785}
\shorttitle{Time evolution of plasma parameters during the rise of a prominence instability}
\shortauthors{Orozco Su\'arez, D\'\i az, Asensio Ramos, and Trujillo Bueno}

\newcommand{\degree}{\ensuremath{^\circ}\/}

\newcommand{\kms}{~km~s$^{-1}$}

\begin{document}

\title{Time evolution of plasma parameters during the rise of a prominence instability}

\author{D.\ Orozco Su\'arez\altaffilmark{1,2}, A.\ J.\ D\'\i az\altaffilmark{3}, A.\ Asensio Ramos\altaffilmark{1,2},  and J.\ Trujillo Bueno\altaffilmark{1,2,4}}

\email{dorozco@iac.es}

\altaffiltext{1}{Instituto de Astrof\'isica de Canarias, E-38205 La Laguna, Tenerife, Spain}
\altaffiltext{2}{Dept.\ Astrof\'isica, Universidad de La Laguna, E-38206 La Laguna, Tenerife, Spain}
\altaffiltext{3}{Departament de F\'\i sica, Universitat de les Illes Balears, E-07122, Palma de Mallorca, Spain,}
\altaffiltext{4}{Consejo Superior de Investigaciones Cient\'ificas, Spain}

\begin{abstract}
We present high-spatial resolution spectropolarimetric observations of a quiescent hedgerow prominence taken
in the \ion{He}{1}~1083.0~nm triplet. The observation consisted of a time series in sit-and-stare mode of
$\sim$~36 minutes of duration. The spectrograph's slit crossed the prominence body and we recorded the time evolution of individual vertical threads. Eventually, we observed the development of a dark Rayleigh-Taylor plume that propagated upward with a velocity, projected onto the plane of the sky, of 17\kms. Interestingly, the plume apex collided with the prominence threads pushing them aside. We inferred Doppler shifts, Doppler widths, and magnetic field strength variations by interpreting the \ion{He}{1} Stokes profiles with the HAZEL code. The Doppler shifts show that clusters of threads move coherently while individual threads have oscillatory patterns. Regarding the plume we found strong red-shifts ($\sim$9-12\kms) and large Doppler widths ($\sim$10\kms) at the plume apex when it passed through the prominence body and before it disintegrated. We associate the red-shifts with perspective effects while the Doppler widths are more likely due to an increase in the local temperature. No local variations of the magnetic field strength associated with the passage of the plume were found; this leads us to conclude that the plumes are no more magnetized than the surroundings. Finally, we  found that some of the threads oscillations are locally damped, what allowed us to apply prominence seismology techniques to infer additional prominence physical parameters. 
\end{abstract}

\keywords{Sun: chromosphere --- Sun: filaments, prominences}

  \section{Introduction}
  \label{sec1}

Quiescent hedgerow prominences are seen as sheet-like plasma structures standing vertically above the solar
surface. They are characterized by the presence of thin, vertically oriented and highly dynamic columns of plasma (hereafter threads) supported against gravity 
\citep[e.g.,][]{2008ApJ...676L..89B,2008ApJ...689L..73C}. They also show plumes that eventually rise from below the prominence and propagate upward with speeds of $\sim$~15\kms\/ \citep{2010ApJ...716.1288B}. These plumes have been recently associated with the magnetic Rayleigh-Taylor (R-T) instability that explains how hot plasma and magnetic flux can be transported upwards through the prominence (e.g.,
\citealt{2005Natur.434..478I,2012ApJ...746..120H}). If the highly dynamic plasma is coupled with the magnetic field, the latter might show local variations, at scales comparable to or
smaller than the typical sizes of the prominence threads. However, there are no direct observational constraints
on magnetic properties of the fine-scale structures seen in hedgerow prominences or on the effects produced
by upward plumes. So far, we only have moderate resolution information about the global magnetic structuring and about
line-of-sight velocities (LOS) and velocities perpendicular to the LOS thought
time-slice techniques using high-spatial resolution filtergrams. (e.g.,
\citealt{1983SoPh...89....3A,1994SoPh..154..231B,2003ApJ...598L..67C,1983SoPh...83..135L,2005SoPh..226..239L,2009ApJ...704..870L,1976ApJ...210L.111L,1983A&A...119..197M,1981SoPh...69..301M,2006ApJ...642..554M,2012ApJ...761L..25O,2013hsa7.conf..786O,1988A&A...197..281S,2010A&A...514A..68S,2013arXiv1309.1568S,1979SoPh...61...39V,1998Natur.396..440Z}).

Here, we report high-resolution full spectropolarimetric measurements taken in the \ion{He}{1}~1083.0~nm triplet, that recorded the temporal evolution of quiescent prominence threads and the passing of a plume generated by a R-T instability through the spectrograph slit.

\begin{figure}[!t]
\begin{center}
\resizebox{\hsize}{!}{\includegraphics{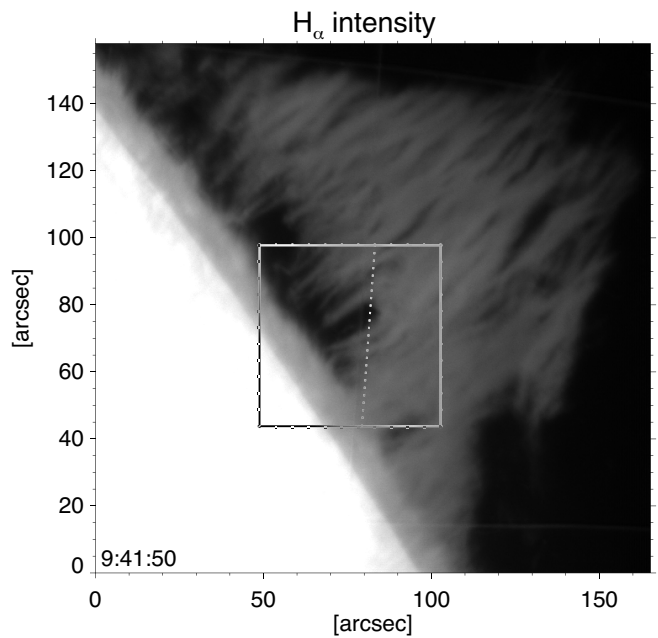}}
\caption{Observed quiescent solar prominence as seen with the H$_\alpha$ slit-jaw camera. The box outlines the area shown in figure 2 that contains a prominence plume. The dotted line represents the TIP-II spectrograph slit, which forms an angle of $\sim$45\degree\/ with the solar limb direction. The observing time (in UT) is shown at the bottom-left.}
\label{fig1}
\end{center}
\end{figure}

\section{Observations, data analysis, and results}
\label{sec2}

\begin{figure*}[!t]
\begin{center}
\resizebox{\hsize}{!}{\includegraphics{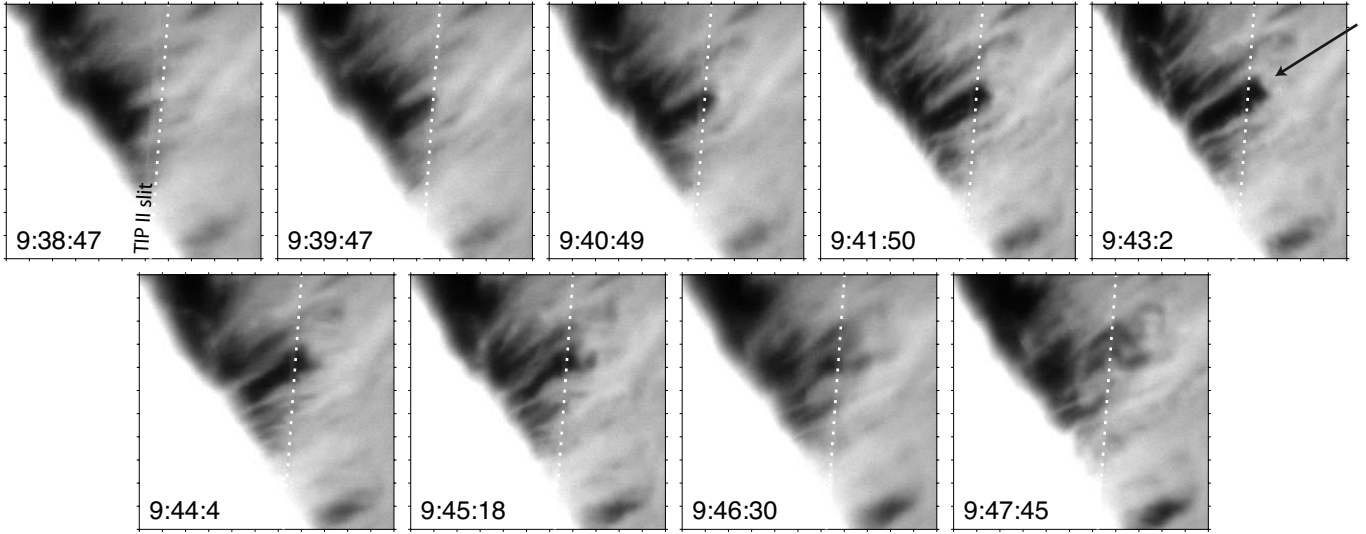}}
\caption{Evolution of the plume as seen in the H$_\alpha$ slit-jaw camera. As in Fig.~1 the dotted line represents the TIP-II slit. The plume is clearly seen around the center of the panels (arrows) and evolves from the left to the right. The observing time (in UT) is shown in each panel.}
\label{fig1}
\end{center}
\end{figure*}

\begin{figure*}[!t]
\begin{center}
\resizebox{\hsize}{!}{\includegraphics{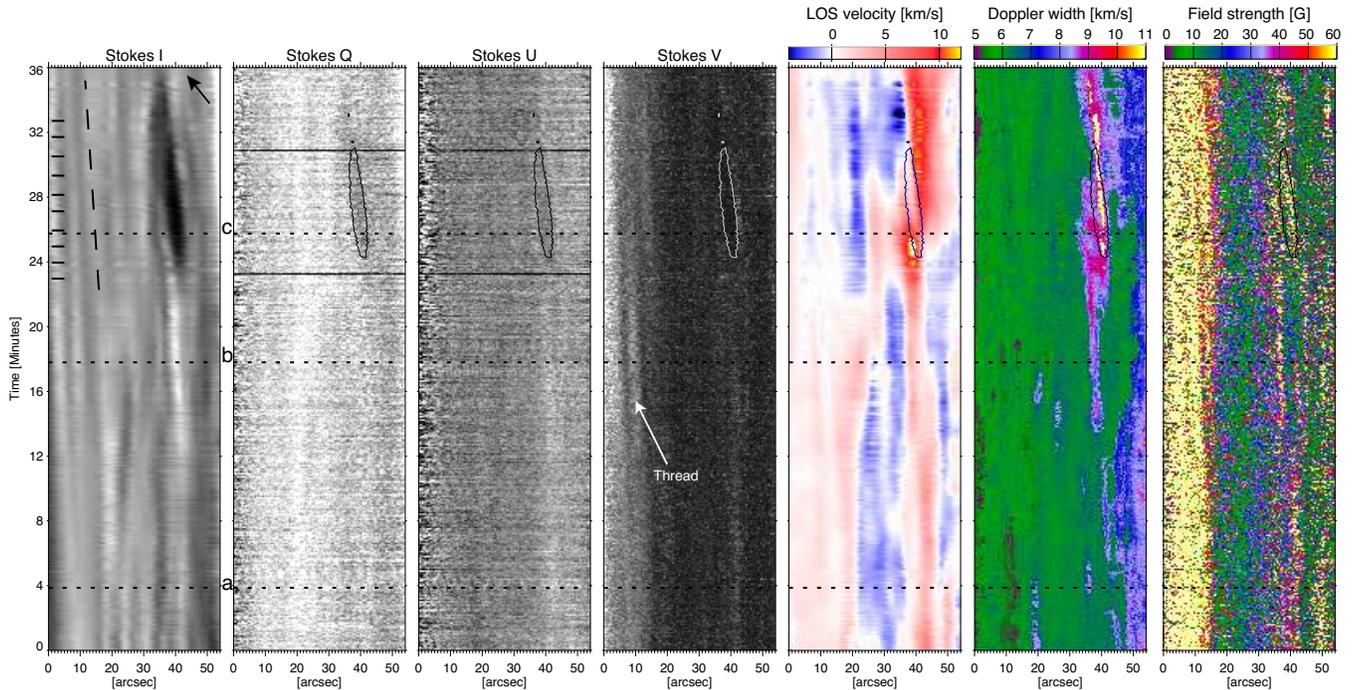}}
\caption{First panel shows the temporal variation (along the Y-axis) of the \ion{He}{1} 1083.0~nm line peak intensity. The X-axis indicates the position along to the TIP-II slit. The arrow indicates the direction from which the prominence plume crosses the slit. The solar limb is at the right side. The horizontal dotted lines represent the cuts presented in Fig.~4 and the vertical dashed line the cut in LOS velocity shown in Fig.~5. The horizontal marks in the left panel indicate the times corresponding to the snapshots displayed in Fig.~2. The rest of the panels show the Stokes Q, U, $|$V$|$ wavelength integrated maps, and the LOS velocity, Doppler width, and field strength resulting from the inversion of the observed Stokes profiles. The contour plot represents the plume. The time evolution of the observed Stokes I profiles is shown in the movie available in the electronic edition of the {\it Astrophysical Journal}.}
\label{fig3}
\end{center}
\end{figure*}

A quiescent hedgerow prominence (see Fig.~1) located in the northwest solar limb was observed on 2 September 2012 with the
Tenerife Infrared Polarimeter (TIP-II; \citealt{2007ASPC..368..611C}) installed of the German Vacuum Tower
Telescope at the Observatorio del Teide (Tenerife, Spain). The TIP-II instrument measured
the four Stokes parameters of the \ion{He}{1}~1083.0~nm triplet with a cadence of 6 seconds (365 slit
images), corresponding to an effective exposure time of 1.25~s per polarization state. The slit was
80\arcsec\/ long with a pixel sampling of 0$.\!\!^{\prime\prime}$17 (rebined to 0$.\!\!^{\prime\prime}$51) and
a width of 0$.\!\!^{\prime\prime}$5. The spectral sampling was 1.1~pm which was reduced to 3.3~pm to increase
the signal-to-noise ratio. The data were processed following the usual procedure including dark current,
flat-field, fringes correction, and polarimetric calibration.  The spatial resolution is close to
1$^{\prime\prime}$. The absolute velocity calibration was done using as reference the averaged spectrum of the Si~{\small
I}~1082.70~nm photospheric line and an atmospheric water vapor line at 1083.21~nm. Then, we corrected the solar and terrestrial rotation and the
relative shifts between the Sun and Earth orbits, after determining the heliographic position and the apparent
height of the solar structure. The heliographic position was estimated using data obtained with the Solar
Terrestrial Relations Observatory (STEREO-B) Extreme UltraViolet Imager \citep{2008SSRv..136....5K}, where the
prominence is seen as a dark absorption feature against the solar disk.

The observed prominence can be identified in the H$_\alpha$ slit-jaw images (see Fig.~1) as an extended body with a foot at the right side. The good seeing conditions during the observing time allow us to distinguish fine scale structures in the prominence body. They are clearly oriented perpendicularly to the solar limb direction as columns of plasma. We have identified these structures as prominence threads. Below the threads (but above the limb) there is a dark cavity (known as prominence bubble) where an instability develops, generating an up-flow plume. The time evolution of the instability can be seen in Fig.~2. A dark cavity below the prominence body is visible at 9:38:47 UT. Then, an instability takes place and a plume develops and rises throughout the prominence from 9:39:47 UT to 9:44:4 UT. The development of the plume seems to destroy the prominence cavity because some of the threads located at the plume sides fall toward the limb at the same time the plume rises. Finally, at 9:45:18 UT, the plume becomes eventually unstable and looses its shape. The disappearance of the plume distorts the prominence thread pattern. Using the H$_\alpha$ slit-jaw images we calculated the rising velocity of the plume,
$\sim17$~\kms, and its lifetime, $\sim$~10 minutes, values that are in line with previous estimations by \cite{2008ApJ...676L..89B}. The plume was best observed in the \ion{Fe}{14} 211 \AA\/ bandpass of the Solar Dynamics Observatory Atmospheric Imaging Assembly  (AIA; \citealt{2012SoPh..275...17L}). Recent analyses of AIA data suggest that the plumes generated by the R-T instability in prominences, contain plasma at temperatures much larger than the surroundings \citep{2011Natur.472..197B}.

Figure~\ref{fig3} shows maps representing the evolution of the \ion{He}{1} 1083.0~nm
triplet peak intensity (associated to its red blended component) and of the Stokes Q, U, $|$V$|$ wavelength integrated signals. The vertical structures seen in the peak intensity map represent the evolution of a cross-section (cutting angle of about 45\degr\/ with respect to the thread)  of individual threads. The thread width, measured as the full-width-at-half-maximum, is about $\sim$2~Mm. The seeing induced jittering is also visible as a horizontal stripe pattern. 
Some of the threads show up in the polarization maps (see arrow).  In the data, 75\% of the pixels show polarization signals above five times their noise level.
 Interestingly, the absolute value of the wavelength integrated Stokes V signals
are greater in the left side than in the right side (closer to the limb). In the peak intensity panel, the
incursion of the plume in the TIP-II slit can be clearly seen 24 minutes after the observation started. The
plume crosses the image of the TIP-II slit diagonally, forming an angle of about 45\degree\/ with the slit and going from
the right side (limb) towards the left side of the map (as indicated by the arrow). 
Although the plume appears black in intensity, there is still detectable Stokes I signals above the noise level, in contrast to the polarization signals which are (if any) below the noise. At the same time the plume appears, there is a bright thread ($\sim$40\arcsec\/ on the slit axis) that seems to be pushed to the left. The evolution of the Stokes I profiles (a movie is available in the electronic edition of the {\it Astrophysical Journal}) shows how the plume pushes a thread aside. It is also evident the presence of strong Doppler shifts in the Stokes I profile  coinciding with the pass of the plume through the slit. 

\begin{figure*}[!t]
\begin{center}
\resizebox{0.8\hsize}{!}{\includegraphics{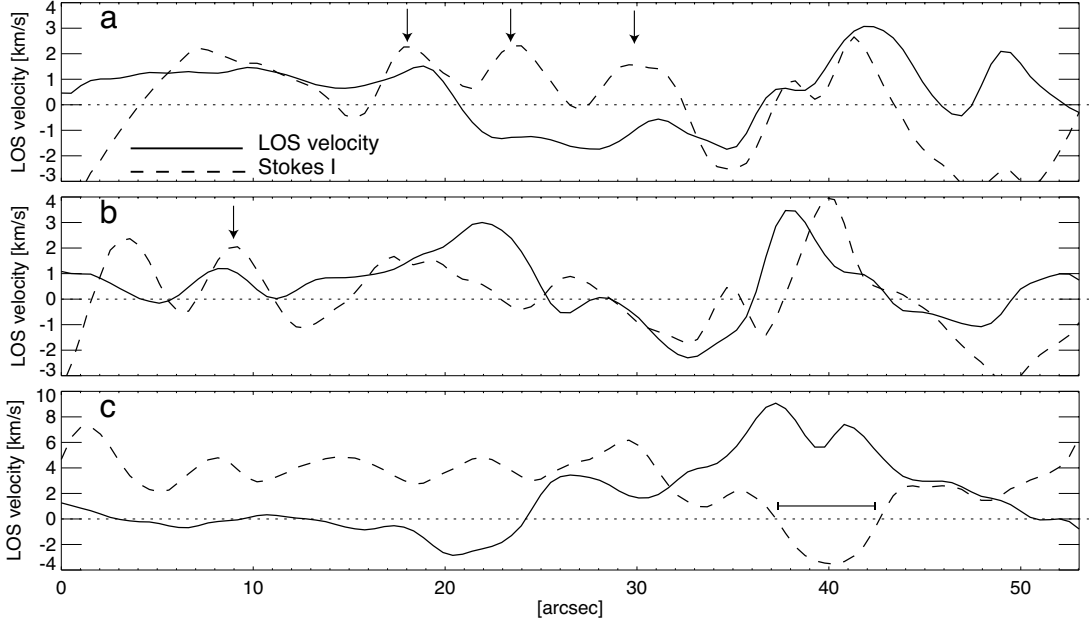}}
\caption{Variation of the LOS velocity (solid) and the peak intensity of the red component of the \ion{He}{1}~1083.0~nm triplet (dashed) along the slit. The a, b, and c letters indicate the location of the cuts in Fig.~3. The vertical arrows pinpoint the location of some of the  threads. The horizontal line in the bottom panel represents the location of the plume.}
\label{fig4}
\end{center}
\end{figure*}

We have analyzed the observed \ion{He}{1}~1083.0~nm triplet polarization signals with the HAZEL  (from HAnle and ZEeman Light, \citealt{2008ApJ...683..542A}) code.  
HAZEL takes into account the Zeeman effect, anisotropic radiation pumping, and the Hanle effect, key physical processes that generate and/or modify circular and linear polarization signals in the \ion{He}{1}~1083.0~nm multiplet \citep{2007ApJ...655..642T}. 
In the inversion we assume a constant-property slab and seven free parameters: the optical depth of the line, the line damping, the Doppler width $\Delta\mathrm{v}_\mathrm{D}$,  the Doppler shift $\mathrm{V}_\mathrm{LOS}$, and the three components of the magnetic field vector. 
We need to give as input the height above the limb and the angle that forms the LOS direction with the local vertical. 
These two quantities help us determine the degree of anisotropy of the incident radiation field and have been determined using the STEREO-B data. 

The right-most panels in Fig.~\ref{fig3} represent the inferred Doppler shifts, thermal velocities, and magnetic field strengths. The Doppler shifts show a prominence Doppler pattern that is not associated with single threads but rather affect clusters of threads as if they were moving in phase (see also \citealt{2010A&A...514A..68S}). 
The pattern changes during the 36 minutes of our observation but not enough to sample large scale oscillations, if they exist in the observed prominence.
The Doppler shifts are at most 5\kms, in absolute value.
At the location where the plume crosses the slit there is a strong redshift signal of $\sim$9-12~\kms\/
(position $\sim$40\arcsec\/ along the slit), slightly before the plume starts to be visible in the slit, at $\sim$~24~min. The strong redshift coincides in location with a thread that is at the same time pushed to the left side by the plume. 
During the passage of the plume through the slit, we also recorded strong redshifts but this time
at the right side of the panel, i.e., between the plume and the prominence cavity above the limb. The strong
redshifts persist after the plume has crossed the slit.

The inferred Doppler widths deviate from the mean value of 6\kms, 6 minutes before the plume appears. Then they show values up to 11~\kms\/ within the plume, and values larger than average, about 8~\kms, after the passage of the plume. On the other hand, the magnetic field strength (with $B\approx15$G around the plume) does not change significantly  
during the rise of the plume. Interestingly, there is a clear field strength gradient, decreasing from left to right.
The inversions indicate that the field vector is 92\degr$\pm$2\degr\/ inclined with respect the solar vertical, with a mean azimuth of 106\degr$\pm$18\degr, and does not show either local nor global spatial variations. This suggests that the vertical columns visible in the He and H$_\alpha$ images in the prominence body are in fact made up by pilling up unresolved small-scale horizontal plasma threads\footnote{Here threads refer to the horizontal fine-scale structures typically seen in high-resolution observations of filaments.}, most likely in magnetic field dips.

\section{Thread Doppler oscillations and prominence seismology}

In Fig.~\ref{fig4} we represent the peak intensity and $\mathrm{V}_\mathrm{LOS}$ spatial variations at three different
time instances a, b, and c (see Fig.~3). The Stokes I peak intensity shows bumps that can be identified as the cross-section of the vertical threads (see arrows) . There are also bumps in the Doppler shifts that, interestingly, are not co-spatial with the bumps seen in the Stokes I peak intensity (see, for instance, the thread indicated by the arrow in panel b, which almost has a correspondence in the Doppler velocity). After a cross-correlation analysis, we conclude that there is no correlation between
the peak intensity and the Doppler shifts pattern. In panel c, the Doppler shifts are considerable larger
than the average in the plume region (horizontal line at $\sim$40\arcsec), reaching values of
8-10\kms\/ at the sides of the plume and of 6\kms\/ at the central part.

If we concentrate on the Doppler shifts associated to single threads, we find that each of them seem to have their own Doppler shift pattern. Interestingly, we have identified the presence of rapidly damped oscillations in some of them, without hints of intensity variations. These Doppler oscillations can be interpreted as transverse kink modes of oscillations, which are nearly incompressible modes that do not produce density or temperature variations and thus modify slightly the emitted intensity (for reviews see \citealt{2012LRSP....9....2A,2010SSRv..151..333M}). The Doppler shift behaviour of each of the threads and the decaying oscillations can be seen in the movie available in the electronic edition of the {\em Astrophysical Journal}. 

By adopting the interpretation that damping is produced by resonant absorption \citep{2008ApJ...682L.141A}, the presence of damped oscillations allows us to perform seismology studies in individual\footnote{We are neglecting the presence of more than one oscillating thread and field line curvature since these effect may be of second order \citep{2009ApJ...693.1601S}.} threads, as proposed by \cite{2011SSRv..158..169A}. Following \cite{2010ApJ...722.1778S}, the oscillation period $P$ and the damping time $\tau$ for a cylindrical prominence thread with a dense core of length $w$, radius $R$ and transition layer of radial extension $l$ (where the density varies linearly from the prominence value $\rho_\mathrm{p}$ to the coronal surrounding value $\rho_\mathrm{c}$) are given by
\begin{equation}
P=\frac{\pi L}{c_\mathrm{Ap}} \sqrt{\frac{w}{L} \left(1-\frac{w}{L} \right)
\frac{1+\rho_\mathrm{c}/\rho_\mathrm{p}}{2}} \approx \frac{\pi}{c_\mathrm{Ap}} \sqrt{\frac{wL}{2}},
\label{per_formula}
\end{equation}
\begin{equation}
\frac{\tau}{P}=\frac{4}{\pi^2} \frac{R}{l}
\frac{\rho_\mathrm{p}+\rho_\mathrm{c}}{\rho_\mathrm{p}-\rho_\mathrm{c}} \approx \frac{4 R}{\pi^2 l},
\label{damp_formula}
\end{equation}
where $c_\mathrm{Ap}=B_0/(\mu \rho_\mathrm{p})^{1/2}$ is the Alfv\'en speed inside the thread and $L$ is the
length of the magnetic field tube in which the thread is embedded. We have assumed that
$\rho_\mathrm{p} \gg \rho_\mathrm{c}$ and $w \ll L$, and that $w/L$ has little influence on the damping ratio \citep{2011A&A...533A..60A} for obtaining the simplified formulae of Eqn.~(1) and (2).

The oscillation period $P$ and the damping time $\tau$ can be inferred by fitting an
oscillatory signal with an exponential decay and a linear trend of the form
\begin{eqnarray}
V_\mathrm{LOS}^\mathrm{fit} (t) = A_0 e^{-(t-t_0)/A_1} \, \cos\left[A_2 (t-t_0) + A_3\right]+A_4(t-t_0)+A_5,
\label{adjust}
\end{eqnarray}
with $A_i$ ($i \in [0,5]$) being the coefficients to be fitted by a least
square algorithm. For the analysis, we have chosen a thread in which this type of oscillation might be present. For example, the thread in the region between 22\arcsec\/ and 34\arcsec\/ and from $t_0\sim$22~min to $t_f\sim$36~min (from 9:35 to 9:49 UT in the movie) displays a clear decaying oscillation (see Fig.~5), while the neighbor regions neither show any
oscillatory pattern nor signs of spatial propagation along the slit. The resulting fit and the coefficients, assuming an error bar for the velocities of $\pm 0.05$ km s$^{-1}$, are shown in Fig.~5. From them we obtain $P= 3.6 \pm 0.2$ min and $\tau=6.73 \pm 0.06$ min. Using  Eq.~\ref{damp_formula} we can determine the ratio $l/R=0.21 \pm 0.01$, which means that, for typical thread tube radius ($50\leq R \leq500$~km, e.g., \citealt{2010ApJ...722.1778S}), the transition layer $l$ is under the achieved spatial resolution.

We can derive other plasma parameters such as $L$, $\rho_\mathrm{p}$, or $w$ using Eq.~\ref{per_formula}, the inferred period and damping time, and the average field strength, $\bar{B}=49\pm 17$~G,  obtained from the interpretation of the Stokes profiles with the HAZEL code in the individual thread. If we want to avoid systematic errors introduced by assuming arbitrary values of any of these quantities, we can only infer the product $(\rho_\mathrm{p}Lw) = 2P^2B_0^2/(\pi^2\mu) = 1.8\times10^5\pm1.5\times10^5$~kg~m$^{-1}$. It would be desirable to analyze two oscillatory events simultaneously since we could obtain an estimation of the length of the supporting magnetic tube from the ratio of different periods \citep{2010ApJ...725.1742D} and then a better estimation of $\rho_\mathrm{p}$ or $w$.

\begin{figure}[!t]
\begin{center}
\resizebox{\hsize}{!}{\includegraphics{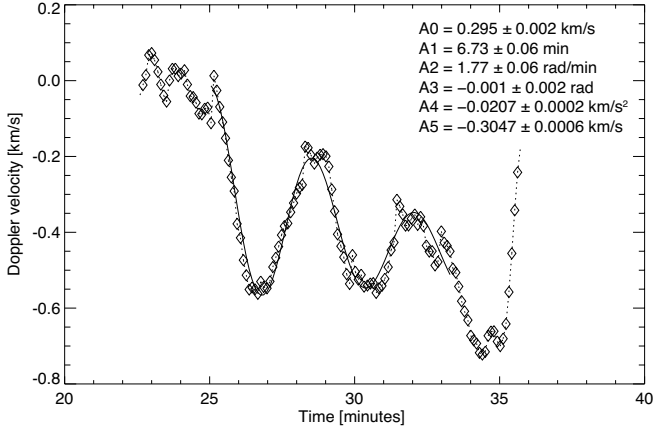}}
\caption{Variation of the Doppler velocities associated with one of the observed prominence threads. The position of the thread is represented with a dashed line in Fig.~3. Diamonds represent the data and the solid line corresponds to a least-squares fit (see sect.~3).}
\label{aj_plot}
\end{center}
\end{figure}

\section{Summary and conclusions}
\label{sec5}

We have presented full spectropolarimetric observations in the \ion{He}{1}~1083.0~nm triplet that
recorded the evolution of quiescent hedgerow prominence threads and the passage of a prominence plume along the
slit and evidence of standing damped oscillations in some of the threads. According to the H$_\alpha$
slit-jaw images, the plume develops at the prominence-cavity interface (probably via the R-T instability) and
rises through the prominence body at about 17~\kms, displacing the thread material horizontally. The rise of
the plume coincides with the disappearance of the underlying cavity. Eventually, the plume disintegrates
distorting the thread pattern. The \ion{He}{1}~1083.0~nm peak intensity shows that the plume apex pushes
the threads aside. Doppler shift measurements indicate that the plasma with which the plume first collides is
strongly redshifted ($\sim$9-12~\kms). Strong redshifts ($<$10~\kms)  are also detected below the plume during
and after the pass of the plume.

If we assume that the plume rises vertically and along the prominence sheet and that the strong red-shifts at
the plume apex correspond to plasma that is being pushed upwards (in other words, we consider that the
velocity component perpendicular to the plane of the sky is negligible) then, the prominence threads and plume
should be slightly inclined about 30\degr\/ from the vertical and away to the observer LOS in order to detect
red-shifts of about 9-12\kms. In this scenario, and since the observed velocity component in the plane of the
sky is about 17~\kms, the plasma at the apex of the plume would be experimenting an upward velocity of
$\sim$~20\kms, which may exceed typical sound speeds in prominences. Other possibility is that the strong
red-shifts are associated with a horizontal (along the LOS) net plasma displacement. The inferred flows would
be then associated with thread material falling along the boundaries of the plume. In this scenario, the plane
containing the prominence threads and plume should also be slightly inclined with respect to the LOS to detect
a net flow.

The Doppler width of the line increases slightly before the plume crosses the slit. It is maximal within the
plume and persists after the plume has disappeared.  The larger Doppler widths within the plume may be due to local temperature enhancements resulting from the injection of cavity's hot plasma into the prominence body. On the other hand, the strong redshifts and the larger Doppler widths after the passage of the
plume could be, in addition, associated with local turbulence generated by the disintegration of plume.

Regarding the field strength, it shows a gradient from left (limb side) to right (corona). The
presence of the plume does not modify the inferred field strengths. Since it is believed that the prominence material is in frozen in conditions, we would expect an increase of magnetic flux at the boundary between the prominence and the plume apex resulting from the squeezing of the field lines in that area. However, we do not detect any sign of flux intensification. Thus, we believe that the plume neither interacts with the magnetic structure of the prominence nor it is more magnetized that the surroundings, which is in line with recent simulation results \citep{khomenko}.

Finally, we found that clusters of threads move together in phase while individual threads show their own (of less amplitude) oscillatory patterns. The presence of oscillations in individual threads have already been detected in solar filaments using the \ion{He}{1}~1083.0~nm multiplet (e.g., \citealt{1991SoPh..132...63Y}). Some of the oscillations are clearly damped and can be associated with transverse kink modes. This allowed us to apply seismology techniques and infer the product $(\rho_\mathrm{p}Lw)$ using the averaged field strength derived from the interpretation of the Stokes profiles with the HAZEL code. Unfortunately, we did not detect other clear decaying oscillations as to compare the results from other threads and determine the length of the field lines, although it is well established that threads tend to oscillate independently and their standing modes are only excited for certain type of perturbations, or oscillate in a plane where little $\mathrm{V}_\mathrm{LOS}$ modulation is produced (\citealt{2002ApJ...580..550D, 2007A&A...469.1135T, 2008ApJ...676..717L}). 

\acknowledgments
The authors, AAR, and AJD acknowledge financial support from the Spanish Ministry of Economy and Competitiveness through the project AYA2010-18029 (Solar Magnetism and Astrophysical Spectropolarimetry), the Ram\'on y Cajal fellowship, and the Spanish Ministry of Education, Culture and Sports through the project CEF11-0012, respectively.


\begin{thebibliography}{}


 \bibitem[Arregui \& Ballester(2011)]{2011SSRv..158..169A} Arregui, I., \& Ballester, J.~L.\ 2011, \ssr, 158, 169 
\bibitem[Arregui et al.(2012)]{2012LRSP....9....2A} Arregui, I., Oliver, R., \& Ballester, J.~L.\ 2012, Living Reviews in Solar Physics, 9, 2
\bibitem[Arregui et al.(2011)]{2011A&A...533A..60A} Arregui, I., Soler, R., Ballester, J.~L., \& Wright, A.~N.\ 2011, \aap, 533, A60 
\bibitem[Arregui et al.(2008)]{2008ApJ...682L.141A} Arregui, I., Terradas, 
J., Oliver, R., \& Ballester, J.~L.\ 2008, \apjl, 682, L141 
\bibitem[Asensio Ramos et al.(2008)]{2008ApJ...683..542A} Asensio Ramos, A., Trujillo Bueno, J., \& Landi Degl'Innocenti, E.\ 2008, \apj, 683, 542
\bibitem[Athay et al.(1983)]{1983SoPh...89....3A} Athay, R.~G., Querfeld,
C.~W., Smartt, R.~N., Landi Degl'Innocenti, E.,
\& Bommier, V.\ 1983, \solphys, 89, 3
\bibitem[Berger et al.(2008)]{2008ApJ...676L..89B} Berger, T.~E., Shine, R.~A., Slater, G.~L., et al.\ 2008, \apjl, 676, L89
\bibitem[Berger et al.(2010)]{2010ApJ...716.1288B} Berger, T.~E., Slater, G., Hurlburt, N., et al.\ 2010, \apj, 716, 1288
\bibitem[Berger et al.(2011)]{2011Natur.472..197B} Berger, T., Testa, P., Hillier, A., et al.\ 2011, \nat, 472, 197
\bibitem[Bommier et al.(1994)]{1994SoPh..154..231B} Bommier, V., Landi Degl'Innocenti, E., Leroy, J.-L., \& Sahal-Brechot, S.\ 1994, \solphys, 154, 231
\bibitem[Casini et al.(2003)]{2003ApJ...598L..67C} Casini, R., L{\'o}pez Ariste, A., Tomczyk, S., \& Lites, B.~W.\ 2003, \apjl, 598, L67
\bibitem[Chae et al.(2008)]{2008ApJ...689L..73C} Chae, J., Ahn, K., Lim, E.-K., Choe, G.~S., \& Sakurai, T.\ 2008, \apjl, 689, L73
\bibitem[Collados et al.(2007)]{2007ASPC..368..611C} Collados, M., Lagg, A., D{\'{\i}}az Garc{\'{\i}} A, J.~J., et al.\ 2007, The Physics of Chromospheric Plasmas, 368, 611
\bibitem[D{\'{\i}}az et al.(2002)]{2002ApJ...580..550D} D{\'{\i}}az, A.~J., Oliver, R., \& Ballester, J.~L.\ 2002, \apj, 580, 550 

\bibitem[D{\'{\i}}az et al.(2010)]{2010ApJ...725.1742D} D{\'{\i}}az, A.~J., Oliver, R., \& Ballester, J.~L.\ 2010, \apj, 725, 1742
\bibitem[Hillier et al.(2012)]{2012ApJ...746..120H} Hillier, A., Berger, T., Isobe, H., \& Shibata, K.\ 2012, \apj, 746, 120 
\bibitem[Isobe et al.(2005)]{2005Natur.434..478I} Isobe, H., Miyagoshi, T., Shibata, K, Yokoyama, T. \ 2005, Nature, 434, 478
\bibitem[Kaiser et al.(2008)]{2008SSRv..136....5K} Kaiser, M.~L., Kucera, T.~A., Davila, J.~M., et al.\ 2008, \ssr, 136, 5
\bibitem[Khomenko et al.(2014)]{2014IAUS..300...90K} Khomenko, E., D{\'{\i}}az, A., de Vicente, A., Collados, M., \& Luna, M.\ 2014, IAU Symposium, 300, 90 Khomenko, Diaz, de Vicente, Collados and Luna, 2014
\bibitem[Lemen et al.(2012)]{2012SoPh..275...17L} Lemen, J.~R., Title, A.~M., Akin, D.~J., et al.\ 2012, \solphys, 275, 17
\bibitem[Leroy et al.(1983)]{1983SoPh...83..135L} Leroy, J.~L., Bommier, V., \& Sahal-Brechot, S.\ 1983, \solphys, 83, 135
\bibitem[Lin et al.(2005)]{2005SoPh..226..239L} Lin, Y., Engvold, O., Rouppe van der Voort, L., Wiik, J.~E., \& Berger, T.~E.\ 2005, \solphys, 226, 239
\bibitem[Lin et al.(2009)]{2009ApJ...704..870L} Lin, Y., Soler, R., Engvold, O., Ballester, J.~L., Langangen, {\O}., Oliver, R.,  and Rouppe van der Voort, L.~H.~M., 2009, \apj, 704, 870
\bibitem[Luna et al.(2008)]{2008ApJ...676..717L} Luna, M., Terradas, J., Oliver, R., Ballester, J.~L., 2008, \apj
   676, 717
\bibitem[Lites et al.(1976)]{1976ApJ...210L.111L} Lites, B.~W., Bruner, E.~C., Jr., Chipman, E.~G., et al.\ 1976, \apjl, 210, L111
\bibitem[Mackay et al.(2010)]{2010SSRv..151..333M} Mackay, D.~H., Karpen, J.~T., Ballester, J.~L., Schmieder, B., \& Aulanier, G.\ 2010, \ssr, 151, 333
\bibitem[Malherbe et al.(1983)]{1983A&A...119..197M} Malherbe, J.~M., Schmieder, B., Ribes, E., \& Mein, P.\ 1983, \aap, 119, 197
\bibitem[Martres et al.(1981)]{1981SoPh...69..301M} Martres, M.-J., Mein, P., Schmieder, B., \& Soru-Escaut, I.\ 1981, \solphys, 69, 301
\bibitem[Merenda et al.(2006)]{2006ApJ...642..554M} Merenda, L., Trujillo Bueno, J., Landi Degl'Innocenti, E., \& Collados, M.\ 2006, \apj, 642, 554
\bibitem[Orozco Su{\'a}rez et al.(2012)]{2012ApJ...761L..25O} Orozco Su{\'a}rez, D., Asensio Ramos, A., \& Trujillo Bueno, J.\ 2012, \apjl, 761, L25
\bibitem[Orozco Su{\'a}rez et al.(2013)]{2013hsa7.conf..786O} Orozco Su{\'a}rez, D., Asensio Ramos, A., \& Trujillo Bueno, J.\ 2013, Highlights of Spanish Astrophysics VII, 786
\bibitem[Schmieder et al.(2010)]{2010A&A...514A..68S} Schmieder, B., Chandra, R., Berlicki, A., \& Mein, P.\ 2010, \aap, 514, A68
\bibitem[Schmieder et al.(1988)]{1988A&A...197..281S} Schmieder, B., Demoulin, P., Poland, A., \& Thompson, B.\ 1988, \aap, 197, 281
\bibitem[Schmieder et al.(2013)]{2013arXiv1309.1568S} Schmieder, B., Kucera, T.~A., Knizhnik, K., et al.\ 2013, arXiv:1309.1568
\bibitem[Soler et al.(2009)]{2009ApJ...693.1601S} Soler, R., Oliver, R., \& Ballester, J.~L.\ 2009, \apj, 693, 1601
\bibitem[Soler et al.(2010)]{2010ApJ...722.1778S} Soler, R., Arregui, I., Oliver, R., \& Ballester, J.~L.\ 2010, \apj, 722, 1778
\bibitem[Terradas et al.(2007)]{2007A&A...469.1135T} Terradas, J., Andries, J., \& Goossens, M. \ 2007 \aap,
   469, 1135.
\bibitem[Trujillo Bueno \& Asensio Ramos(2007)]{2007ApJ...655..642T} Trujillo Bueno, J., \& Asensio Ramos, A.\ 2007, \apj, 655, 642 
\bibitem[Vial et al.(1979)]{1979SoPh...61...39V} Vial, J.~C., Gouttebroze, P., Artzner, G., \& Lemaire, P.\ 1979, \solphys, 61, 39
\bibitem[Yi et al.(1991)]{1991SoPh..132...63Y} Yi, Z., Engvold, O., \& Keil, S.~L.\ 1991, \solphys, 132, 63 
\bibitem[Zirker et al.(1998)]{1998Natur.396..440Z} Zirker, J.~B., Engvold, O., \& Martin, S.~F.\ 1998, \nat, 396, 440

\end{thebibliography}
\end{document}